\documentclass[prb,twocolumn,showpacs,amsmath,amssymb,superscriptaddress]{revtex4}
\usepackage{graphicx}
\usepackage{dcolumn}
\usepackage{bm}
\usepackage{epsfig}

\begin{document}
\title{Magnetic transition and spin dynamics in the triangular Heisenberg
  antiferromagnet $\alpha$-KCrO$_{2}$}
\author{F. Xiao}
\affiliation{Durham University, Department of Physics, South Road,
  Durham, DH1 3LE, UK}
\author{T. Lancaster}
\affiliation{Durham University, Department of Physics, South Road,
  Durham, DH1 3LE, UK}
\author{P.J. Baker}
\author{F.L. Pratt}
\affiliation{ISIS Pulsed Neutron and Muon Facility, STFC Rutherford
  Appleton Laboratory, Harwell Oxford, Didcot, OX11 OQX, UK}
\author{S.J. Blundell}
\affiliation{University of Oxford, Department of Physics, Clarendon
  Laboratory, Parks Road, Oxford, OX1 3PU, UK}
\author{J.S. M\"{o}ller}
\affiliation{University of Oxford, Department of Physics, Clarendon
  Laboratory, Parks Road, Oxford, OX1 3PU, UK}
\author{ N.Z. Ali}
\affiliation{Max-Planck-Institut f\"{u}r Festk\"{o}rperforschung,
  Heisenbergstr.\ 1, 70569 Stuttgart, Germany}
\author{M. Jansen}
\affiliation{Max-Planck-Institut f\"{u}r Festk\"{o}rperforschung,
  Heisenbergstr.\ 1, 70569 Stuttgart, Germany}

\date{\today}

\begin{abstract}
We present the results of muon-spin relaxation measurements on the
triangular lattice Heisenberg antiferromagnet $\alpha$-KCrO$_{2}$. 
We observe sharp changes in behaviour at an ordering temperature of 
$T_{\mathrm{c}}=23$~K, with an additional broad feature in the
muon-spin relaxation rate evident at $T=13$~K, both of which
correspond to features in the magnetic contribution to the heat
capacity. 
This behaviour is
distinct from both the Li- and Na- containing members of the series.
These data may be qualitatively described with the established
theoretical
predictions for the underlying spin system.

\end{abstract}
\pacs{75.10.Jm, 76.75.+i, 75.40.Cx}
\maketitle

The geometrical frustration of antiferromagnetic (AFM) interactions continues to be important as
a route to creating exotic quantum mechanical ground states including various flavours of quantum
spin liquid \cite{sachdev,wen}. Frustration most obviously occurs for
antiferromagnetically coupled spins on 
a triangular lattice but, until recently, there
have been relatively few examples of materials well described by a
model of Heisenberg spins on such a lattice. 
In this context the series
$A$CrO$_{2}$ (where $A=$ Li, Na, K) is of interest as its members 
comprise well-decoupled, highly ideal, triangular planes containing
isotropic spins. 
While the members of the series with $A=$~Li
\cite{delmas,soubeyroux,angelov,kadowaki,mazin,olariu_li,ikedo} and Na
\cite{delmas, soubeyroux, elliston, angelov, olariu_na} have been  
well studied, the material 
KCrO$_{2}$, which has the best-separated triangular layers and might be expected to best 
approximate the model, has previously proved difficult to stabilise
chemically
and has been the subject of less experimental work\cite{delmas,soubeyroux,scheld}.
However, it was recently demonstrated\cite{ali}
that KCrO$_{2}$ may be reliably synthesized in two different
polymorphs, the $\alpha$- and $\beta$-phases, allowing the
opportunity for renewed study. 
Muon-spin relaxation ($\mu^{+}$SR) has proven particularly useful in probing both the
NaCrO$_2$\cite{olariu_na} and LiCrO$_2$\cite{olariu_li,ikedo} materials, 
suggesting unusual spin relaxation spectra in both systems below their
respective short- or long-range ordering
temperatures,  with an exotic fluctuating regime reported well below
the short-range ordering temperature $T_{\mathrm{c}}$ in NaCrO$_2$, which has better separated layers than LiCrO$_2$. 
Here we present the results of muon-spin relaxation measurements on
$\alpha$-phase KCrO$_{2}$ ($\alpha$-KCrO$_2$).  We find that the fluctuation spectrum in this
compound, while showing features that are superficially similar to
both the Na- and Li- containing materials, appears to be distinct and
unusual. Moreover, although we find that $\alpha$-KCrO$_{2}$ may be 
described qualitatively within the framework of two different theoretical
approaches, it does not seem to admit a quantitative description
consistent with either of them.

The structure of the $A$CrO$_2$ series comprises well separated triangular planes
of  $S=3/2$ Cr$^{3+}$ ions stacking in an $ABCABC$ sequence
with $R\bar{3}m$ symmetry.
The separation of the triangular sheets ($1/3$ of the lattice constant $c$) is found to vary from
 4.81~\AA\ for $A=$ Li to 5.96~\AA\ for $A=$ K, leading us to expect that the potassium-containing material 
will possess the most magnetically isolated two-dimensional layers.
Electron spin resonance measurements on LiCrO$_2$ and NaCrO$_2$
\cite{elliston,angelov} indicate very small single-ion anisotropy,
suggesting a strong Heisenberg character for the spins. 
The structural and magnetic parameters of the members of the series are summarized in Table~\ref{tab:parameters}.
Below $T_{\mathrm{N}}=62$~K, LiCrO$_2$ undergoes a
transition to a phase of long-range magnetic order (LRO),  adopting a
120$^{\circ}$ magnetic
structure \cite{kadowaki} with a suggestion of AFM
coupling between layers \cite{mazin}.
Short-range magnetic order (SRO) has been proposed to occur in NaCrO$_2$ below $T_{\mathrm{c}}=41$~K. 
 No magnetic Bragg peaks (indicative of LRO) were observed in
$\alpha$-KCrO$_{2}$ down to 5~K, but SRO was originally suggested to occur at
$T_{\mathrm{c}}=26$~K on the strength of a diffuse neutron scattering peak \cite{soubeyroux}. The transition is also confirmed by the more recent 
heat capacity measurement where a peak is observed at 23 K and by magnetic susceptibility\cite{ali} (where the extracted Fisher heat capacity~\cite{fisher} shows a relatively sharp peak at 24 K). 
Previous $\mu^{+}$SR measurements on the Li-\cite{olariu_li} and Na-containing \cite{olariu_na}  materials
revealed quite different behaviour in the muon-spin relaxation rate
close to the respective ordering temperatures.
In LiCrO$_{2}$ there is a sharp peak in the relaxation rate 
at  $0.9 T_{\mathrm{N}}$. This contrasts with the behaviour of the
more two-dimensional NaCrO$_{2}$
where the relaxation rate shows a far broader peak, with a maximum at
$0.75 T_{\mathrm{c}}$. This unusual behavior was justified by considering the nature of the fluctuations in 
triangular lattice antiferromagnets. It was suggested that NaCrO$_{2}$ has an 
exotic, extended fluctuating regime, with a slow freeze-out of fluctuations below 41~K rather than a conventional ordering transition. It was
speculated that this may relate to $Z_{2}$ topological defects that have been predicted to occur in the spectrum of the
triangular lattice. In contrast, the state of affairs in the more three-dimensional LiCrO$_{2}$ leads to something closer to the expected 
behavior of a relaxation rate peaked at $T_{\mathrm{N}}$. 
\begin{table}
   
     \begin{ruledtabular}
\centering
\renewcommand{\arraystretch}{1.5}
\begin{tabular}{>{\centering\arraybackslash}m{0.5in} >{\centering\arraybackslash}m{0.8in} >{\centering\arraybackslash}m{0.8in} >{\centering\arraybackslash}m{0.8in} }
 &LiCrO$_2$ &NaCrO$_2$ &KCrO$_2$\\\hline
 $a$ (\AA) &2.898 &2.975 &3.042 \\\hline
 $c$ (\AA) &14.423 &15.968 &17.888 \\\hline
 $J$ (K)\footnote{Ref~\onlinecite{delmas} uses the Hamiltonian $-2J\sum_{ij}\mathbf{S}_i\cdot\mathbf{S}_j$ and we use the Hamiltonian $-J\sum_{ij}\mathbf{S}_i\cdot\mathbf{S}_j$, therefore the values of $J$ reported in Ref~\onlinecite{delmas} are doubled in the table for consistency.} &78 &40 &24\\\hline
 $\theta_{\mathrm{CW}}$ (K) &-620\cite{soubeyroux}, -700~\cite{delmas} &-290~\cite{delmas} &-160~\cite{delmas}, -220~\cite{ali}\\\hline
 $T_\mathrm{c}$ (K) &62~\cite{soubeyroux, alexander} &41~\cite{soubeyroux, olariu_na} &26~\cite{soubeyroux}, 23~\cite{ali}\\
 \end{tabular}
     \end{ruledtabular}
\caption{Structural and magnetic parameters for compounds in the $A$CrO$_2$ series ($A=$~Li, Na, K). The cell parameters $a$ and $c$ are  from Ref.~\onlinecite{soubeyroux} and the exchange strengths $J$ are from Ref.~\onlinecite{delmas}. $\theta_\mathrm{CW}$ is the Curie-Weiss constant derived from the magnetic susceptiblity fit.}\label{tab:parameters}
\end{table}

In a $\mu^{+}$SR experiment \cite{steve} spin polarized muons
are implanted into the sample. The quantity of interest
is the angular asymmetry of the decay positrons, $A(t)$, which is proportional to the spin polarization of the muon ensemble.
Zero-field muon-spin relaxation (ZF $\mu^{+}$SR) measurements
were made on a polycrystalline
sample of $\alpha$-KCrO$_{2}$ using the EMU spectrometer at the ISIS
facility, UK. 
A sample of $\alpha$-phase KCrO$_{2}$ was prepared as previously
described \cite{ali}.
It was packed, under glove-box conditions,
inside a Ti-foil packet (foil thickness $25~\mu$m) and sealed in an
air-tight Ti sample holder, which was mounted inside a $^{4}$He cryostat.
The measurements described here were made in the temperature range
$1.5\leq T \leq 200$~K. After the measurements were completed the sample was heated above 500~K in an attempt to detect a transition to the $\beta$ phase, which we were unable to stabilize. 

\begin{figure}[h]
\centering
\includegraphics[width=\linewidth]{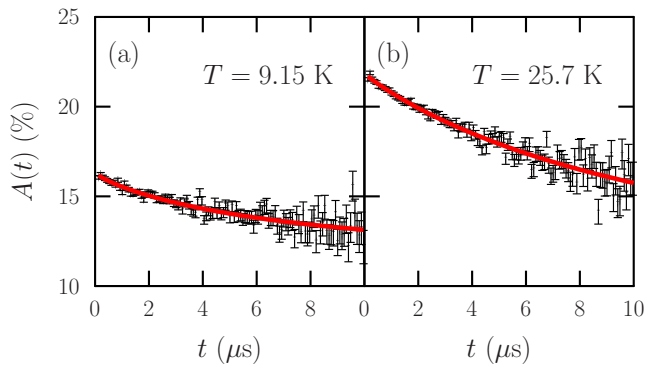}
\caption{\label{fig:muon-spectra} (Color online.) Example $A(t)$ spectra of poly-crystalline 
$\alpha$-KCrO$_2$ at (a) 9.15~K and (b) 25.7~K. Solid lines (red) represent the fit to Eq.~(\ref{eqn:fit-model}).}
\end{figure}

Example ZF $\mu^{+}$SR spectra are shown in Fig.~\ref{fig:muon-spectra}.
Below 23~K we observe a loss of asymmetry in the
signal with a transition region $20\lesssim T\lesssim 23$~K. This is due to the time resolution limit of the ISIS muon
pulse, which prevents the observation of features with rate $\gtrsim
10$~MHz. 
The loss of asymmetry demonstrates the presence of large,
slowly fluctuating moments (see below) which depolarize those muon
spin components perpendicular to the local magnetic field (expected to be 2/3 of the total in a powder sample~\cite{hayano}). Although it is not possible to 
tell whether this results from a state of static LRO or
SRO on the basis of these measurements, the absence of magnetic Bragg peaks and the fact that
the layer separation is greater than for NaCrO$_{2}$ where SRO is thought
to occur, leads us to tentatively assign $T_{\mathrm{c}}$ as a short-range
ordering temperature. 

\begin{figure}[h]
\centering
\includegraphics[width=\linewidth]{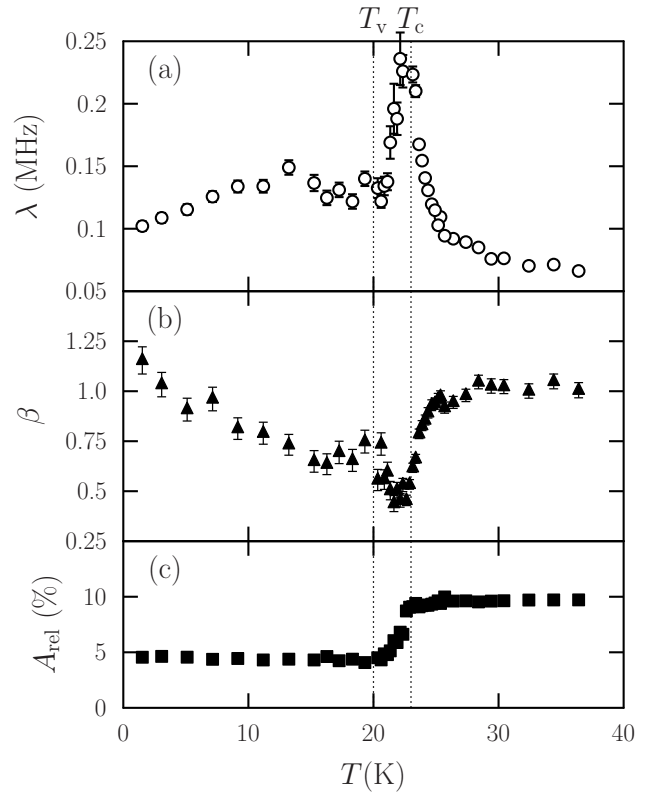}
\caption{\label{fig:fit-parameters}
Evolution of the parameters from Eq.~(\ref{eqn:fit-model}) with
temperature. Sharp features are observed at $22.6(5)$~K
along with a broad maximum in $\lambda$ centred on 13~K. $T_\mathrm{c}$ is identified from heat capacity peak and $T_\mathrm{v}$ is defined in the discussion (see main text).}
\end{figure}

In order to compare to previous muon measurements on similar
triangular materials \cite{olariu_na,olariu_li,zhao},
the muon spectra were fitted to a model  of stretched exponential
decay with a background contribution:
\begin{equation}
A(t) = A_\mathrm{rel}e^{-(\lambda t)^\beta}+A_\mathrm{bg},
\label{eqn:fit-model}
\end{equation}
where $A_\mathrm{rel}$ represents the amplitude of the relaxing
component, $\lambda$ is the relaxation rate and 
$A_\mathrm{bg}$ accounts for the constant background contribution from muons stopping in the sample holder or cryostat tails. 
The values of $\lambda$, $\beta$ and $A_\mathrm{rel}$ obtained from
the fit are plotted against temperature  in Fig.~\ref{fig:fit-parameters}. 
All three parameters show distinct changes at $22.6(5)$~K, in agreement with heat capacity measurements \cite{ali} where a sharp peak was observed
at $T_\mathrm{c}=23$~K. In addition to the sharp peak in $\lambda$, whose rapid decrease levels off below 20 K, the relaxation rate also shows a broad maximum near 13~K, similar to that observed in NaCrO$_2$~\cite{olariu_na}.

 
 
The relaxation rate $\lambda$ in Eq.~(\ref{eqn:fit-model}) is expected
to vary with the width of the local magnetic field distribution
$\Delta$ and correlation time $\tau$ according to
$\lambda\propto\Delta^{2}\tau$,  and therefore the peak in 
$\lambda$ is suggestive of a transition to a regime of large, slowly fluctuating moments as the temperature approaches $T_\mathrm{c}$. 

\begin{figure}[h]
\centering
\includegraphics[width=\linewidth]{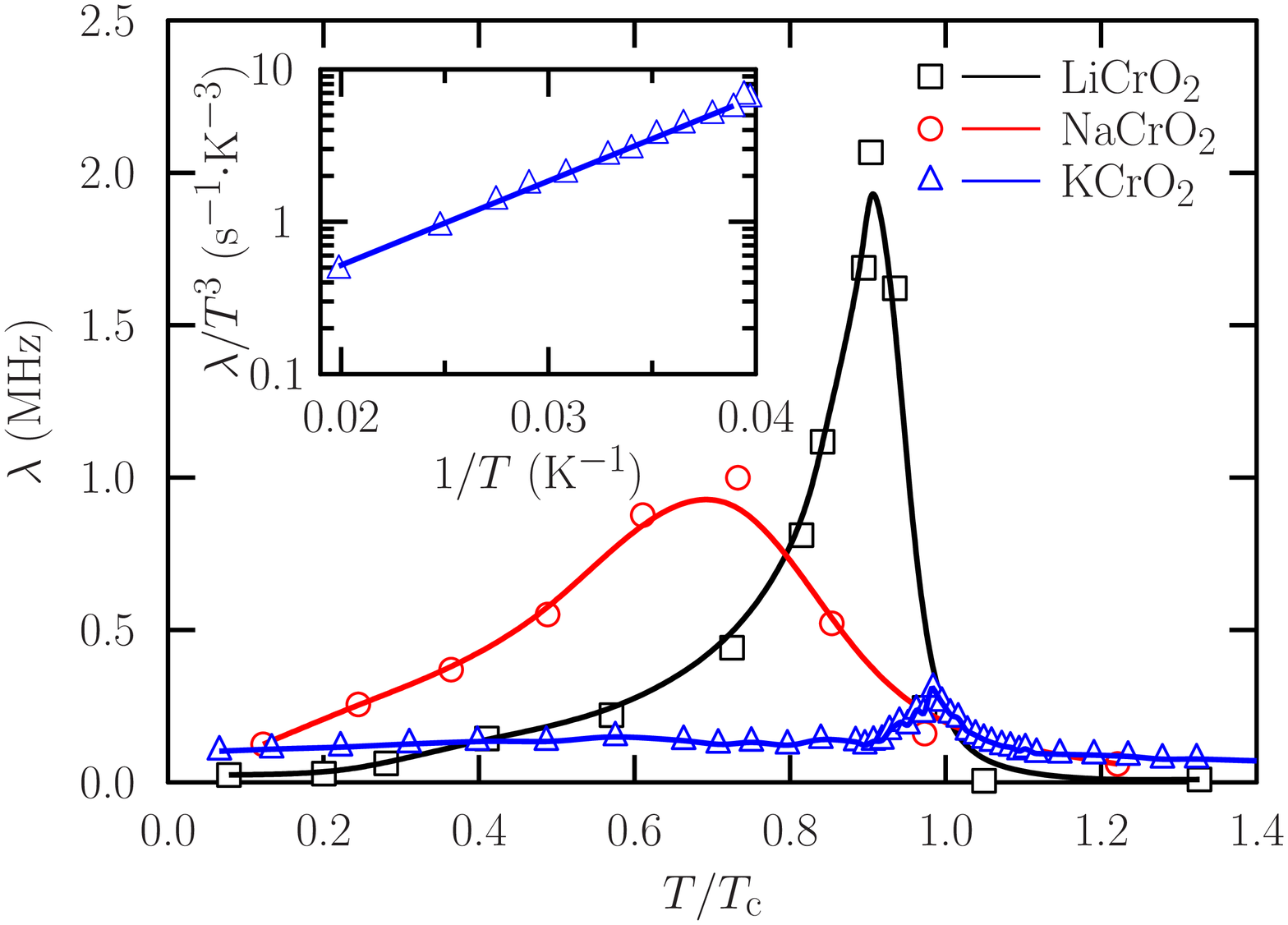}
\caption{\label{fig:usr-comparison}
(Color online.) Comparison of relaxation rates for materials in the series
$A$CrO$_{2}$
with $A=$ Li, Na and K. The solid lines are a guide to the eye. The peak of $\alpha$-KCrO$_2$ appears to be located at a value slightly less than 1, this is due to the experimental uncertainty (we take $T_\mathrm{c}=23~$K from the heat capacity peak for $\alpha$-KCrO$_2$). Inset: Dependence of $\lambda/T^3$ on inverse temperature $1/T$ for $\alpha$-KCrO$_2$. The solid line represents the fit to Eq.~(\ref{eqn:css-model}) for $T>T_{\mathrm{c}}$ and the slope in this semi-log plot corresponds to $T_0/\ln10$.}
\end{figure}

The muon-spin relaxation rates for all three compounds in the $A$CrO$_{2}$
family are plotted in Fig.~\ref{fig:usr-comparison} with normalized $x$-axis. 
For $\alpha$-KCrO$_{2}$, the maximum in $\lambda$ occurs at $22.6(5)$~K and we take $T_\mathrm{c}=23$~K from the heat capacity peak.
The relaxation rate for $\alpha$-KCrO$_{2}$ is noticeably smaller than the Li-
and Na- compounds, pointing to significantly smaller moments or to shorter
fluctuation times.  
For LiCrO$_2$, the sharp peak (corresponding to 3D LRO) occurs a few Kelvin below $T_{\mathrm{N}}$. 
NaCrO$_2$ shows no features close to $T_{\mathrm{c}}$, but instead we see a broad peak below the
critical temperature with the maximum centred on 0.75$T_{\mathrm{c}}$. 
Our measurements on $\alpha$-KCrO$_2$, superficially seem to show a
combination of both features: 
not only a sharp peak corresponding to an ordering transition very close to $T_{\mathrm{c}}=23$~K, but also a very broad shoulder with a maximum centred on $0.57T_{\mathrm{c}}$. 

\begin{figure}[h]
\centering
\includegraphics[width=\linewidth]{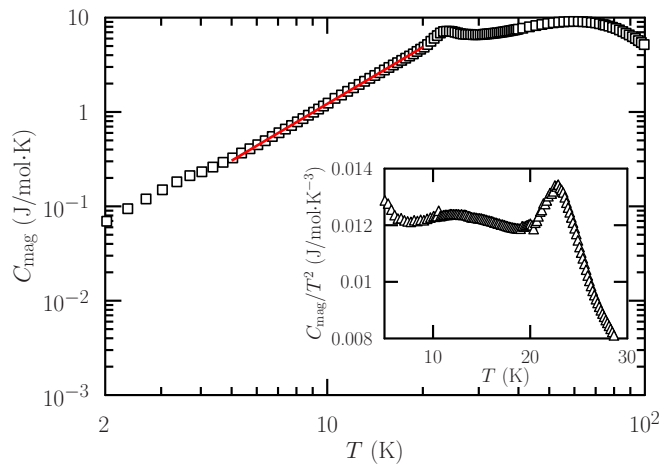}
\caption{\label{fig:specific-heat} 
The magnetic specific heat ($C_\mathrm{mag}$) of $\alpha$-KCrO$_{2}$. The solid line is a $T^2$ fit to the data between 5~K and 20~K. The inset shows $C_\mathrm{mag}/T^2$ as a function of temperature.}
\end{figure}

It is illuminating to compare the heat capacity results for
$\alpha$-KCrO$_{2}$ \cite{ali} with our muon measurements.
The phonon component of $C_{\mathrm{p}}$ was obtained by fitting the
high-$T$ data 
to a Debye model with $\theta_{\mathrm{D}}=552$~K. 
The lattice contribution was then subtracted from the total 
$C_{\mathrm{p}}$ so only the magnetic heat capacity $C_{\mathrm{mag}}$
is plotted in Fig.~\ref{fig:specific-heat}.
The low-$T$ region of $C_{\mathrm{mag}}$ ($5\leq T \leq 20$~K) exhibits
a $T^{2}$  temperature dependence (solid line in Fig.~\ref{fig:specific-heat}),
which implies linearly dispersing 2D excitations. This form is also
observed for triangular materials showing SRO such as NiGa$_2$S$_4$ \cite{zhao}. 
In contrast, $C_{\mathrm{mag}}$ for LiCrO$_2$ (which shows LRO) was found to have a $T^{3}$  dependence~\cite{alexander} below $T_\mathrm{c}$, consistent with  a recent spin-wave theory calculation~\cite{du}. 
It is also interesting to note that below 3~K  $C_{\mathrm{mag}}$
suggests $T^2$  behaviour but with a different scaling prefactor.
In Fig.~\ref{fig:specific-heat} (inset) the evolution of $C_\mathrm{mag}/T^{2}$ is shown. In addition to the sharp peak at $T_{\mathrm{c}}$,  a broad
shoulder is present below $T_{\mathrm{c}}$ with a maximum around 13~K, 
similar to that observed in the muon relaxation rate. No such shoulder is reported for NaCrO$_2$.

The results of our $\mu^+$SR measurements may be compared against different
theoretical descriptions of the triangular lattice AFM Heisenberg model. 
Above $T_{\mathrm{c}}$, the spin-wave theory developed by Chubukov,
Sachdev, and Senthil (CSS)~\cite{css} 
predicts that, for $T \ll 2\pi\rho_{\mathrm{s}}$, the muon relaxation
rate follows~\cite{zhao}
\begin{equation}
\lambda=G_\mathrm{CSS}\left(\frac{N_0A_0}{\hbar}\right)^2\frac{\hbar}{\rho_s}\left(\frac{T}{T_0}\right)^3\exp(T_0/T),
\label{eqn:css-model}
\end{equation}
where $N_0A_0$ is a renormalized hyperfine coupling constant,
$\rho_{\mathrm{s}}$ is the spin stiffness constant, $G_\mathrm{CSS}$ is a numerical
constant and $T_0=4\pi\rho_{\mathrm{s}}$. 

By fitting the data above
$T_{\mathrm{c}}$ to Eq~(\ref{eqn:css-model}), we are able to obtain 
$T_0$ and $\rho_s$. The fit to the data (Fig.~\ref{fig:usr-comparison} inset) yields $T_0=130(2)$~K and $\rho_s=10.3(2)$~K, which is consistent with the limit $T<2\pi\rho_s$ for the validity of the model. With the value of $\rho_s$, the effective exchange coupling $J$ can then be calculated based on the approximation~\cite{css, lecheminant},
\begin{equation}
\frac{\rho_{\mathrm{s}}}{JS^2}=\frac{1-0.399/2S}{\sqrt{3}},
\label{eqn:spin-wave-J}
\end{equation}
where $S$ is the spin quantum number ($3/2$ for $\alpha$-KCrO$_2$). The spin-wave exchange strength is found to be $J_{\mathrm{sw}}=9.2~$K.
Although the high-$T$ dependence of the relaxation rate is well described by this model,
the exchange constant $J$ is much smaller than that obtained from the
Curie-Weiss susceptibility fit, where $J_{\theta}$ is found to be
29.3~K~\cite{ali} using $|\theta_\mathrm{SW}|=zJS(S+1)/3$ with $z=6$, or the value from high temperature series expansion fit obtained in Ref~\onlinecite{delmas}, where $J=24$~K. 
We therefore find a significant discrepancy between the quantitative values, which is unlike
the case of NiGa$_2$S$_4$ \cite{zhao} where good agreement was obtained for this model. However
 it is worth noting that the qualitative agreement with
 Eq.~(\ref{eqn:css-model}) might suggest that the CSS model captures
 some of the underlying physical behaviour.

Below $T_\mathrm{c}$, the experimental results are compared to another theoretical description of the triangular lattice system which has been
 invoked to describe previous $\mu^{+}$SR results \cite{olariu_na,zhao},
known as the spin-gel picture \cite{kawamura, kawamura_2} . Here it is suggested that vortex excitations and spin freezing provide two length scales which  determine the behaviour: the vortex correlation
length $\xi_{\mathrm{v}}$ 
 and spin-wave correlation length $\xi_{\mathrm{sw}}$. 
The vortex correlation length diverges below a topological
critical temperature $T_{\mathrm{v}}$, where $Z_{2}$ vortex excitations undergo a
binding transition. However,  the spin-wave correlation length remains
finite below this temperature, causing the
overall effective spin correlation length to remain finite also~\cite{kawamura_2}. 
This $T_{\mathrm{v}}$ is predicted to lie slightly below the peak temperature in the heat capacity, $T_\mathrm{peak}$~\cite{kawamura}. Specifically, quantum Monte Carlo simulations~\cite{kawamura} have shown that with this model we should expect $T_{\mathrm{peak}}=0.137\, \theta_{\mathrm{CW}}$ and
$T_{\mathrm{v}}=0.123\,\theta_{\mathrm{CW}}$
for classical Heisenberg triangular AFM lattices, where
$\theta_{\mathrm{CW}}$ is the Curie-Weiss constant extracted from fits
of the magnetic susceptibility.  
This picture may be applied to $\alpha$-KCrO$_{2}$ if we identify 
$T_{\mathrm{peak}}$ with $T_\mathrm{c}=23$~K and $T_{\mathrm{v}}$ with $T=20$~K, below which the heat capacity $C_\mathrm{mag}$ follows a $T^2$ trend due to the dominance of spin-wave excitations and also where the rapid change in muon relaxation rate levels off. Using the relation between $T_\mathrm{peak}$, $T_\mathrm{v}$ and $\theta_{\mathrm{CW}}$, two values of $\theta_{\mathrm{CW}}$ are obtained $\theta_{\mathrm{CW}1}=23/0.137=168$~K and $\theta_{\mathrm{CW}2}=20/0.123=163$~K. (In Ref~\onlinecite{kawamura} $T_\mathrm{v}$ is identified with the rounded shoulder in $\lambda$ for  NaCrO$_2$. If we adopt the same procedure a value of $\theta_{\mathrm{CW}3}=13/0.123=106~$K is derived, which is inconsistent with the value from $T_\mathrm{peak}$.) The calculated $\theta_{\mathrm{CW}}$ values are in reasonable agreement with the $\theta_{\mathrm{CW}}=160~$K measured in Ref~\onlinecite{delmas} but about 25\% smaller than the more recent $\theta_{\mathrm{CW}}=220~$K measurement in Ref~\onlinecite{ali}.  We note that the latter value of $\theta_\mathrm{CW}=220$~K corresponds to the material we measured, where the $\alpha$-phase was successfully isolated. Given this, and the ambiguity in identifying the features in the data corresponding to $T_\mathrm{v}$ it is unclear whether this model is applicable in this case.

In conclusion we have made $\mu^{+}$SR measurements on
$\alpha$-phase KCrO$_{2}$. The material undergoes a transition, most
probably to a region of short-range magnetic order below
$T_{\mathrm{c}}=23$~K and shows evidence for further dynamics below
this temperature with a peak seen in the muon-spin relaxation rate and
a broad shoulder in the magnetic heat capacity at 13~K. Despite the superficial
resemblance to the muon relaxation seen in the Li- and Na- containing
members of the series, the features here appear to be unique. Although the behaviour is qualitatively consistent with two established models of the 2D triangular Heisenberg antiferromagnet, namely the CSS spin-wave theory and the spin-gel picture, a fully consistent quantitative description is not obtained with either model.

This work was carried out at the STFC ISIS Facility. 
We are grateful to STFC for the provision of muon beamtime and to EPSRC (UK) for financial support.

\end{document}